\documentstyle[12pt]{article}

\begin{document}

\vskip 2. truecm
\centerline{\bf Chiral condensate, susceptibilities, critical coupling}
\centerline{\bf and indices in QED$_4$.}
\vskip 2 truecm

\centerline { V.~Azcoiti$^a$, G. Di Carlo$^b$, A. Galante$^{c,b}$,
A.F. Grillo$^d$,  V. Laliena$^a$, C. Piedrafita$^a$}
\vskip 1 truecm
\centerline {\it $^a$ Departamento de F\'\i sica Te\'orica, Facultad
de Ciencias, Universidad de Zaragoza,}
\centerline {\it 50009 Zaragoza (Spain).}
\vskip 0.15 truecm
\centerline {\it $^b$ Istituto Nazionale di Fisica Nucleare,
Laboratori Nazionali di Frascati,}
\centerline {\it P.O.B. 13 - Frascati 00044 (Italy). }
\vskip 0.15 truecm
\centerline {\it $^c$ Dipartimento di Fisica dell'Universit\'a
dell'Aquila, L'Aquila 67100 (Italy)}
\vskip 0.15 truecm
\centerline {\it $^d$ Istituto Nazionale di Fisica Nucleare,
Laboratori Nazionali del Gran Sasso,}
\centerline {\it Assergi (L'Aquila) 67010 (Italy). }
\vskip 3 truecm
\centerline {ABSTRACT}
We measure chiral susceptibilities in the Coulomb phase of noncompact
QED$_4$ in $8^4, 10^4$ and $12^4$ lattices. The
MFA approach allows simulations in the chiral limit which are
therefore free from arbitrary mass extrapolations. Using the
critical couplings extracted
from these calculations, we study the critical behaviour of the chiral
condensate, which we find in disagreement with the predictions of
logarithmically improved scalar Mean Field theory.
\vfill\eject

\section{INTRODUCTION}

The question of the non triviality of QED$_4$ defined at the finite coupling
critical point of the theory is long standing [1-4].
While there is agreement
on the numerical results obtained by various groups, their interpretation
differs markedly. In \cite{TED} good agreement is found fitting the
equation of state
(i.e. critical exponents  and the critical coupling from the numerical
 simulations)
with  the predictions of a logarithmically improved Mean Field scalar
theory, finding, for four flavours, $\beta_c = 0.186(1)$.

The numerical determination of the critical exponents is very
sensitive to the exact value of the critical coupling. In \cite{HOMBRES,NOS}
different simulations, entirely independent, found perfect agreement with
a value of $\beta_c$ larger than $0.186$ and critical exponents
implying non Mean Field behaviour.

In this paper we present a new precise determination of the critical
coupling from the behaviour of the chiral susceptibilities as computed in
the Coulomb phase. Chiral susceptibilities (both longitudinal and
transverse) are well defined in this phase also in the chiral limit,
so that their behaviour can
be used to locate the phase transition. In the MFA approach \cite{MFA},
this can be
done directly in the chiral limit $m_f=0$, thus avoiding arbitrary mass
extrapolations. Such an approach has been discussed in \cite{SUSC}.

The value of the critical coupling so obtained (which
agrees with the one reported in \cite{NOS}) is then used to measure the
critical exponent of the chiral condensate, which is compared with the
expectations from (logarithmically improved) Mean Field theory \cite{SACHA}.

\section{THE CHIRAL SUSCEPTIBILITIES}

In the MFA approach \cite{MFA}
the light fermion dynamics  is expressed through an effective action
\begin{equation}
Z(\beta,m)=\int dE n(E) e^{-6V\beta E -S_f}
\end{equation}
where $S_f=ln\overline{\det} \Delta(E,m)$  is
the average of the fermionic determinant
over gauge configurations at fixed euclidean
energy, and $n(E)$ is the density of states.
A modified Lanczos algorithm \cite{LANCZOS} is used
to obtain all the eigenvalues of the fermionic matrix at $m_f=0$, so that
results at different couplings and masses are trivially related to $m_f=0$
and do not need separate simulations.

In the same way, average of physical operators can be expressed in terms of
microcanonical averages of the corresponding operators:
\begin{equation}
\langle O \rangle_{\beta} \Leftrightarrow \langle O \rangle_E
\end{equation}

We use this approach to compute the average value of
\begin{equation}
O={2\over V} \sum_i  {1\over{\lambda^2_i}}
\end{equation}
which defines both longitudinal and transverse chiral susceptibilities at
$m_f=0$ in the
symmetric phase of the model; this term diverges at the transition in the
thermodynamical limit.
In the chirally broken phase, the
suceptibilities differ, and cannot be defined if not starting at non zero
mass and extrapolating to the massless limit, thus suffering from
ambiguities.
On the contrary, chiral susceptibility in the Coulomb phase is free from
arbitrary mass extrapolations.

We have performed simulations with
staggered fermions on lattices of sizes $L=8,10,12$,
and $n_f=0,2,8$. The results we present here are at
$n_f=4$; $L=10$ ($300-400$ configurations per energy value)
and $L=12$ ($60-120$ configurations).
In Fig. 1
we present the inverse massless susceptibility as function of
Euclidean energy. The inverse susceptibility shows
a clear linear behaviour throughout the whole Coulomb phase. The
fermionic effective action is linear in this phase, this implying that the
whole effect of dynamical fermions amounts merely to a shift in the
coupling $\beta=\beta_{PG}+r$ with respect to the pure gauge theory,
$r$ depending on the number of flavours,
and suggests
that the inverse suceptibility can be parametrized in the symmetric
phase as
\begin{equation}
{\chi}^{-1} = {c\over{(\beta_c+r)}} {(\beta_c - \beta)\over{(\beta+r)}}
\end{equation}
as a function of the coupling (Fig. 2).
The critical coupling $\beta_c$ can be derived directly from the fit of
Fig. 2; alternatively a critical energy can be derived from
the beahviour of the inverse susceptibility versus energy, and then
converted to $\beta_c$ through the knowledge of the effective action and
density of states.

We thus obtain $\beta_c= 0.2025(2),\quad L=10$ $(0.2051(2), \quad L=12$)

\section{THE CHIRAL CONDENSATE}
We use the above derived values of the critical coupling to study the
behaviour of the chiral condesate versus mass at the phase transition.
Mean Field theory {\it a la} NJL predicts \cite{SACHA}
\begin{equation}
\langle \bar \psi \psi \rangle \times
ln{1\over \langle \bar \psi \psi \rangle} \propto m_f
\quad (\beta = \beta_c, m_f \to 0)
\end{equation}

We have tested the above relation in our simulation, in the fermion mass
range $0.01 - 0.06$. Due to the fact that we
compute all the eigenvalues of the fermionic matrix in the chiral limit,
the chiral condensate we derive is essentially known continuosly as
function of $m_f$.

In Fig. 3 we present the best fit of the Mean Field
prediction (Eq. 5) with respect to the data.
Mean Field prediction and data are clearly inconsistent, with
$\chi^2/(d.o.f)=57$.

We have fitted the experimental data with the relation
\begin{equation}
\langle \bar \psi \psi \rangle \propto m^{1/\delta}_f \quad
(\beta=\beta_c,m_f \to 0)
\end{equation}
obtaining very good fits with $\delta=2.89(2), \quad L=10 \quad (2.79(7),
\quad L=10)$ (Fig. 4) in the same mass range.

\section{CONCLUSIONS}

The use of the MFA approach allows a very precise computation of the
massless chiral susceptibility in the symmetric phase of QED$_4$, where,
due to the unbroken chiral symmetry, the interchange of the chiral and
thermodynamical limit is guaranteed. The inverse chiral susceptibility is
to extremely good precision linear in the whole Coulomb phase, this
allowing a very precise determination of the critical energy unaffected by
finite volume effects near the transition.
Through the knowledge of the effective action and
density of states this leads to a very precise determination of the
critical coupling \cite{SUSC}.

We use
this information to study the critical behaviour of the chiral condensate
at the phase transition as a funtion of $m_f$, and compare it with the
expectations with Mean Field theory {\it a la}  NJL. We find definite
disagreement, while the data  are well fitted by a power of the mass, with
an inverse power significantly
smaller than 3. We remark that the scalar field motivated Mean
Field solution advocated by \cite{TED} would imply an effective
exponent {\it larger} than 3, thus {\it a fortiori} excluded by our
analysis.

If,  on the other hand, we take the Mean Field prescription (Eq. 5) and
allow for an arbitrary power of the logarithm,
\begin{equation}
\langle \bar \psi \psi \rangle \times
ln^q{1 \over \langle \bar \psi \psi \rangle} \propto  m_f
\end{equation}
we can find a fit as good as that presented in Fig. 4
with $q=0.255$. We obviously
cannot exclude such a behaviour, at least in the same mass range;
larger interval of masses, towards small ones,
where the predictions of Eqs. 6,7 have to differ,
can only be reached in lattices larger than ours.

\vskip 0.5 truecm

The numerical simulations quoted above have been done using the Transputer
Networks of the Theoretical Group of the Frascati National Laboratories
and the Reconfigurable Transputer Network (RTN),
a 64 Transputers array, of the University of Zaragoza.

This work has been partly supported through a CICYT (Spain) -
INFN (Italy)
collaboration.

\newpage
\vskip 1 truecm
\leftline{\bf Figure captions}
\vskip 1 truecm

\noindent
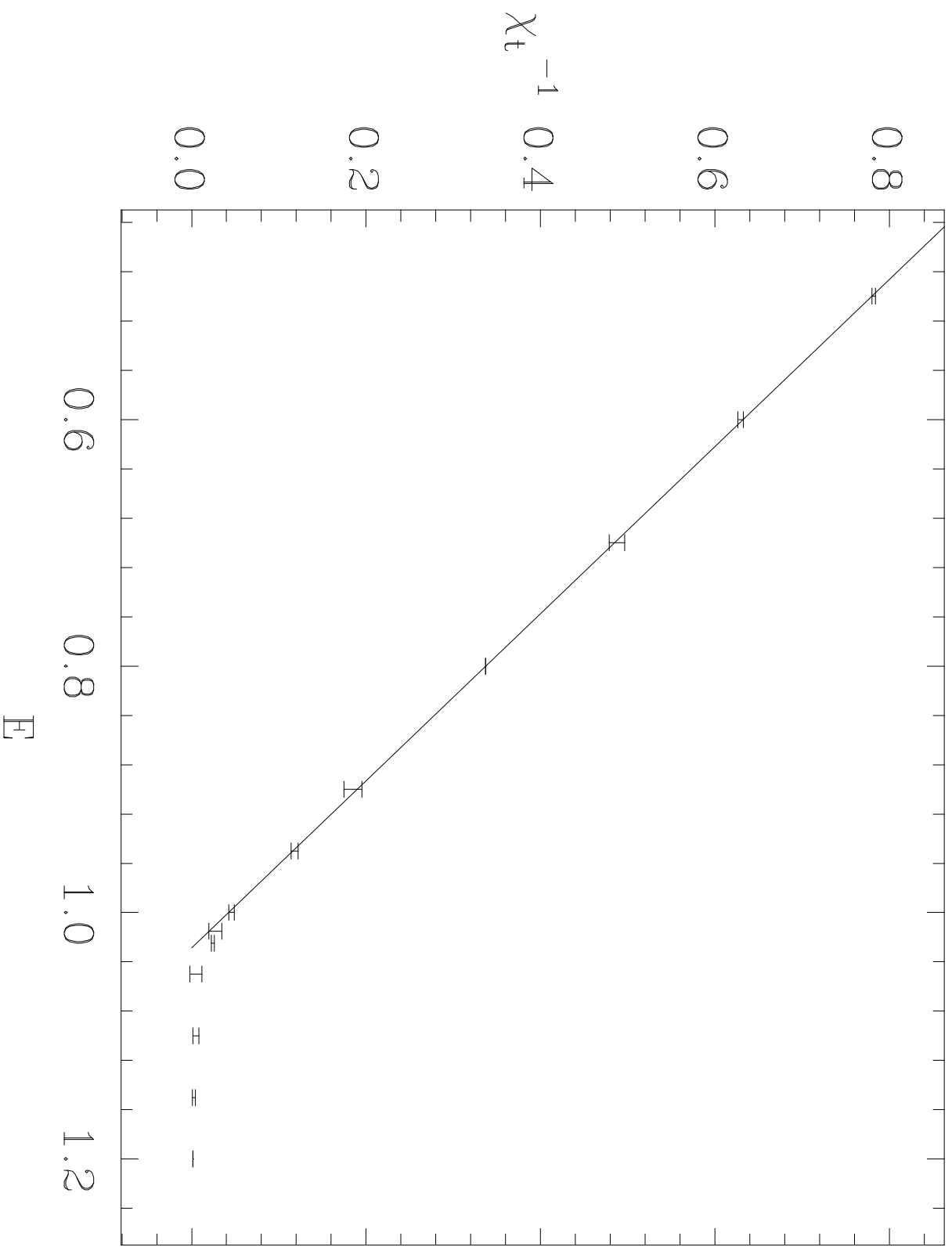
{\bf Figure 1.} Inverse susceptibility as function of
Euclidean Energy, $L=10$.

\noindent
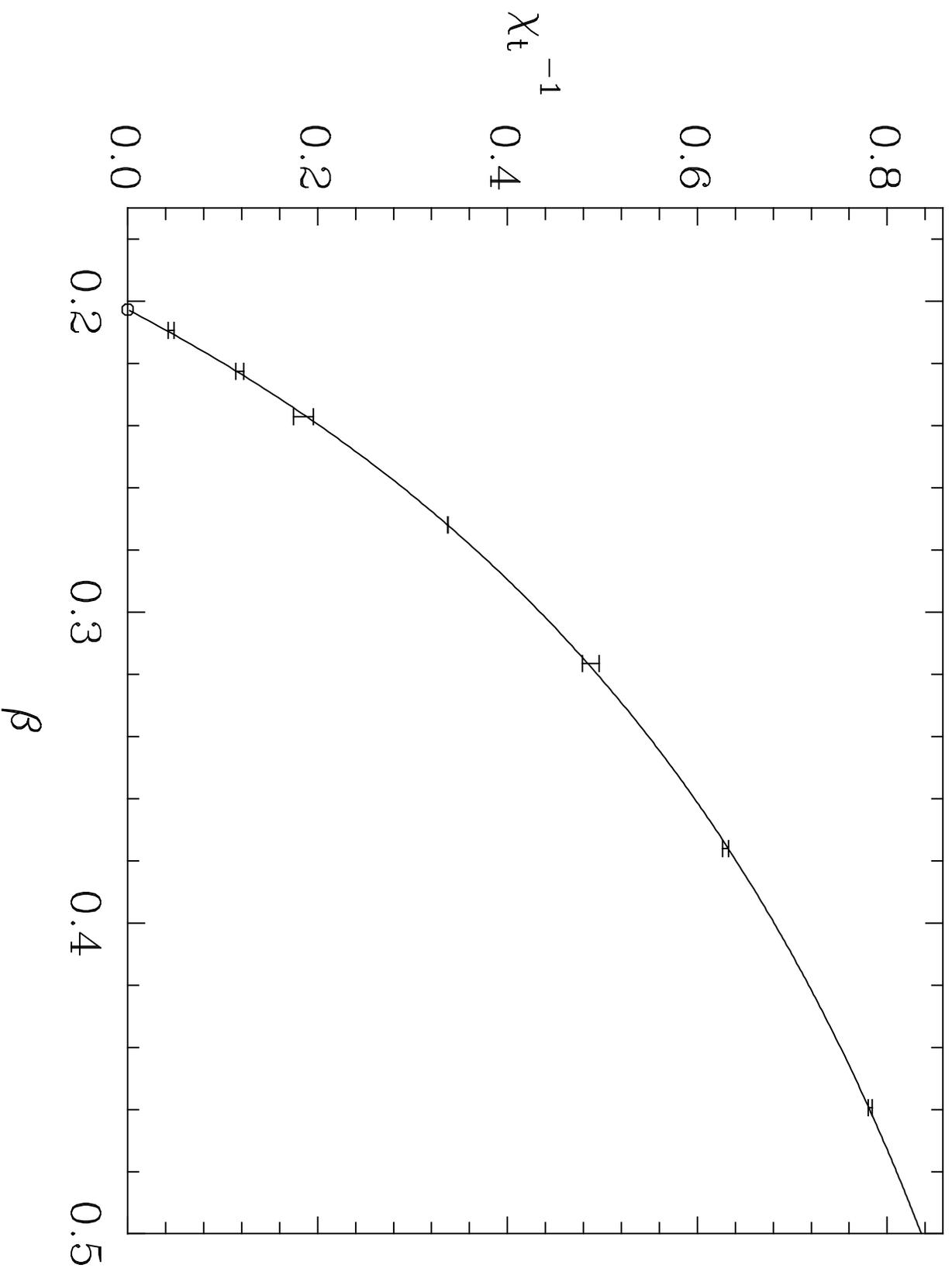
{\bf Figure 2.} Inverse susceptibility vs. $\beta$, $L=10$.
The continuous line is a fit with Eqn. 4.

\noindent
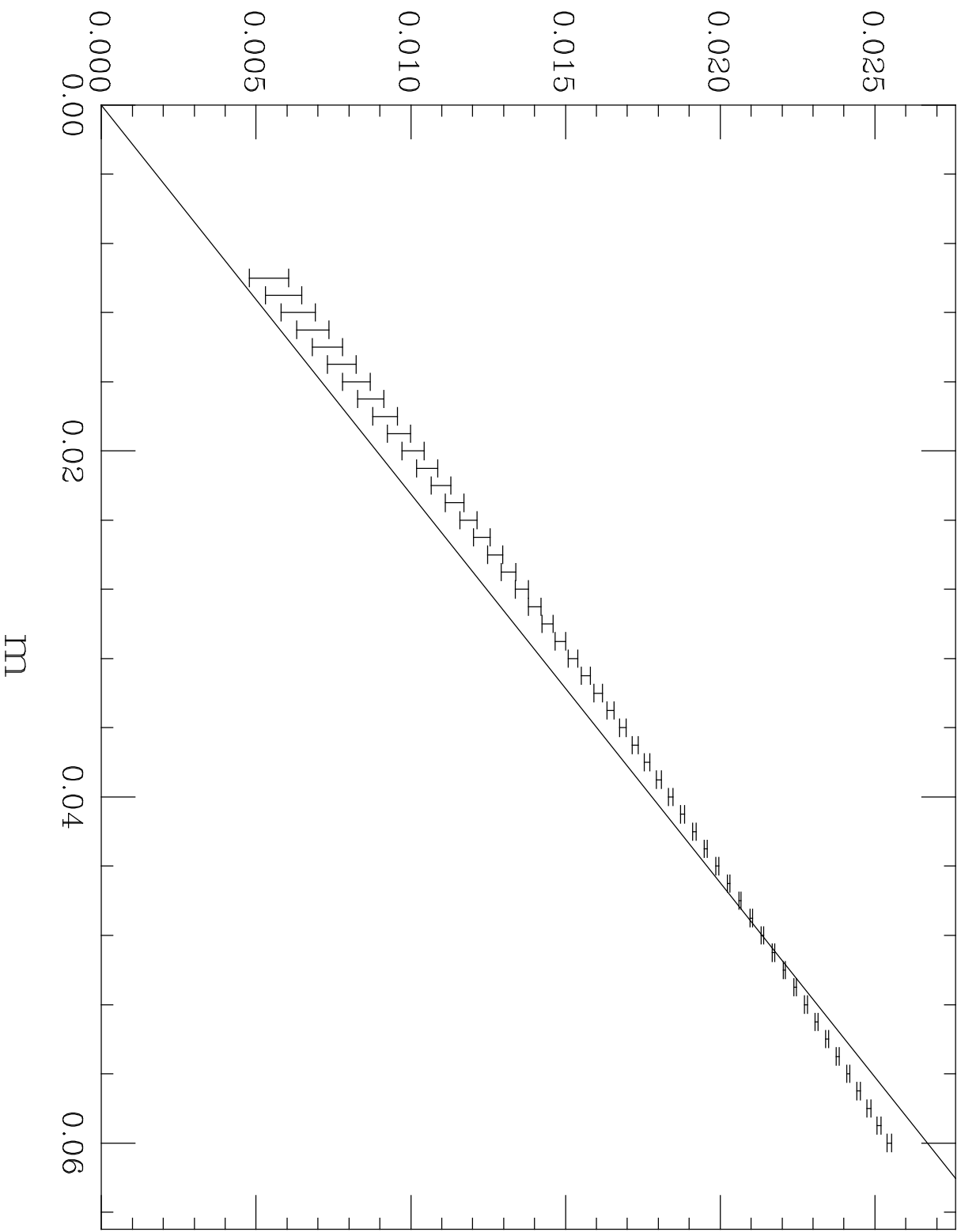
{\bf Figure 3.} Mean Field prediction for the chiral condensate
(continuous line) compared with the data; L=10.

\noindent
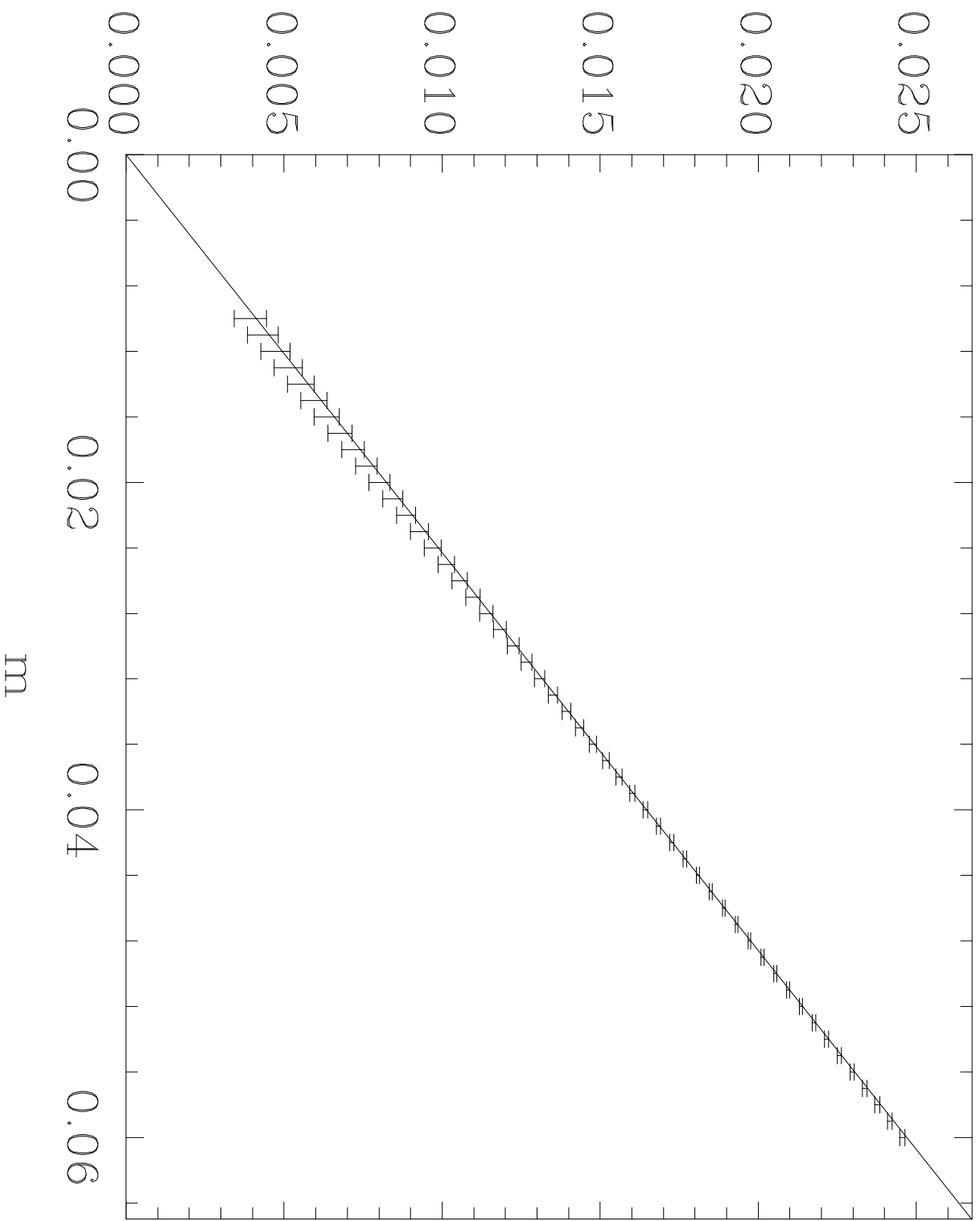
{\bf Figure 4.} $m^{1/\delta}_f$ (continuous line)
compared with the data; L=10.

\end{document}